%% file: paper.tex
\RequirePackage{fix-cm}

\documentclass[smallextended]{svjour3}       % onecolumn (second format)
\smartqed  % flush right qed marks, e.g. at end of proof
\usepackage[pdftex]{graphicx}
\usepackage{cite}
\usepackage{amsmath,amsfonts,bm,amssymb}
\usepackage[caption=false,font=footnotesize]{subfig}
\usepackage{url}
\usepackage{paralist}
\usepackage{tikz}
\usetikzlibrary{shapes.geometric, arrows, fit, positioning}
\usepackage[htt]{hyphenat}
\input{newcommands}
\begin{document}
\title{Modeling Hierarchical Usage Context for Software Exceptions based on
Interaction Data}
\titlerunning{Modeling Hierarchical Usage Context of Software Exceptions}

\author{
    Hui Chen,
    Kostadin Damevski,
    David Shepherd \and
    Nicholas A. Kraft
}
\institute{H. Chen \at 
    Department of Computer and Information Science \\
    Brooklyn College of the City University of New York \\
    Brooklyn, NY 11210, U.S.A.\\
    \email{huichen@ieee.org} 
    \and
    K. Damevski \at 
    Department of Computer Science \\
    Virginia Commonwealth University \\
    Richmond, VA 23284, U.S.A.\\
    \email{damevski@acm.org}
    \and
    D.C. Shepherd \at
    ABB Corporate Research\\
    Raleigh, NC 27606 U.S.A.\\
    \email{david.shepherd@us.abb.com}
    \and
    N.A. Kraft \at
    ABB Corporate Research\\
    Raleigh, NC 27606 U.S.A.\\
    \email{nicholas.a.kraft@us.abb.com}
}

\date{Received: date / Accepted: date}

\maketitle

\input{abstract}

\input{sec_intro}
\input{sec_bg}
\input{sec_nhdp}
\input{sec_eval}

\input{sec_threat}

\input{sec_related}

\input{sec_conclusion}

\begin{acknowledgements}
    The authors would like to thank the RobotStudio team at ABB Inc for providing
    the interaction dataset and responding to the survey. The authors are also
    grateful to the anonymous reviewers' constructive comments. 
\end{acknowledgements}

\bibliographystyle{spmpsci}      % basic style, author-year citations
\bibliography{paper}

\end{document}

%% file: newcommands.tex
\newcommand{\cardinality}[1]{\left\vert{#1}\right\vert}
\newcommand{\set}[1]{\mathbb{#1}}
\newcommand{\vect}[1]{\boldsymbol{\mathbf{#1}}}

\newcommand{\code}{\texttt}

\newcommand{\toctree}[1]{\mathcal{T}}

\makeatletter
\newcommand{\distas}[1]{\mathbin{\overset{#1}{\kern\z@ \sim}}}%
\newsavebox{\mybox}\newsavebox{\mysim}
\newcommand{\distras}[1]{%
  \savebox{\mybox}{\hbox{\kern3pt$\scriptstyle#1$\kern3pt}}%
  \savebox{\mysim}{\hbox{$\sim$}}%
  \mathbin{\overset{#1}{\kern\z@\resizebox{\wd\mybox}{\ht\mysim}{$\sim$}}}%
}
\makeatother

\mathchardef\mhyphen="2D
\newcommand\mheldout{{held\mhyphen out}}
\newcommand\mobs{{obs}}

%% file: abstract.tex
\begin{abstract}

    Traces of user interactions with a software system, captured in production,
    are commonly used as an input source for user experience testing. In this
    paper, we present an alternative use, introducing a novel approach of
    modeling user interaction traces enriched with another type of data gathered
    in production - software fault reports consisting of software exceptions and
    stack traces. The model described in this paper aims to improve developers'
    comprehension of the circumstances surrounding a specific software exception
    and can highlight specific user behaviors that lead to a high frequency of
    software faults.

    Modeling the combination of interaction traces and software crash reports to
    form an interpretable and useful model is challenging due to the complexity
    and variance in the combined data source. Therefore, we propose a
    probabilistic unsupervised learning approach, adapting the Nested
    Hierarchical Dirichlet Process, which is a Bayesian non-parametric
    hierarchical topic model originally applied to natural language data. This
    model infers a tree of topics, each of whom describes a set of commonly
    co-occurring commands and exceptions. The topic tree can be interpreted
    hierarchically to aid in categorizing the numerous types of exceptions and
    interactions. We apply the proposed approach to large scale datasets
    collected from the ABB RobotStudio software application, and evaluate it
    both numerically and with a small survey of the RobotStudio developers.

    \keywords{
        Stack Trace, Crash Report, Software Exception, Software Interaction
        Trace, Hierarchical Topic Model
    }

\end{abstract}

%% file: sec_intro.tex
\section{Introduction}

% talk about bug prioritization and talk about how interaction traces have
% rarely been considered for this purpose.

%
% stack-trace is important data debugging, in particular, for software
% in production.
%
% it is important to understand bugs, in particular, the context of
% the bugs, a characteristics of bugs

Continuous monitoring of deployed software usage is now a standard approach in
industry. Developers leverage usage data to discover and correct faults,
performance bottlenecks, or inefficient user interface design.  This practice
has led to a debugging methodology called ``debugging in the large'', a
postmortem analysis of large amount of usage data to recognize patterns of
bugs~\cite{Han:2012:PDL:2337223.2337241, glerum2009debugging}. For instance,
Arnold et al.\ use application stack traces to group processes exhibiting
similar behavior called ``process equivalence classes'', and identify what
differentiate these classes with the aim to discover the root cause of the bugs
associated with the stack traces~\cite{arnold2007stack}. Han et al.\ clusters
stack traces and recognize patterns of stack traces to discover impactful
performance bugs~\cite{Han:2012:PDL:2337223.2337241}.

%Following the above, we begin our work with the following observation.
Software-as-a-service applications often gather monitoring data at the service
host, while user-installed client software collects relevant traces (or logs)
periodically at the user's machines and transferred them from users'
machines to a server. The granularity and format of the collected data (e.g.,
whether the format of the data is a raw/log form or as a set of derivative
metrics) depend on the specific application and deployment. Two types of data
commonly collected via monitoring include {\em software exceptions}, containing
a stack traces from software faults that occur in production, and {\em
interaction traces}, containing details of user interactions with the
software's interface.

By utilizing datasets that contain both of these two types of data, we can provide 
a novel perspective on interpreting frequently occurring stack traces resulting from software exceptions by
modeling them in concert with the user interactions with which they co-occur.
Our approach probabilistically represents stack traces and their interaction context
for the purpose of increasing developer understanding of specific software
faults and the contexts in which they appear. Over time, this understanding can
help developers to reproduce exceptions, to prioritize software crash reports
based on their user impact, or to identify specific user behaviors that tend to
trigger failures.  Existing works attempt to empirically characterize software
crash reports in application domains like operating systems, networking
software, and open source software
applications~\cite{Yin:2010:TUB:1823844.1823849, Chou:2001:ESO:502059.502042,
Li:2006:TCE:1181309.1181314, Lu:2008:LMC:1353535.1346323}, but none have used
interaction traces containing stack traces for the purpose of fault characterization 
debugging.

%% why it is innovative to use interaction or behavior data to understand bugs
%Much software requires human interaction, such as, any software with user
%interface. During users' interaction with a piece of software, a software fault
%may happen, and it leads to an exception, followed by a software crash.  During
%the process, many messages are written to log traces. These messages often
%include stack traces and messages representing the interactions between the
%users and the software.  Capturing the interactions between the user and her
%software and the stack traces, we can analyze the interactions and establish in
%which context the exception may occur, useful information to the developers.
%Noting that developers' interaction traces with IDE have not been used for
%understanding software bugs, we posit that a hierarchical analysis can
%effectively lead to understanding the context of exceptions, and potentially
%software bugs.

% Why hierarchical models?
Interaction traces can be challenging to analyze. First, the logged
interactions are typically low-level, corresponding to most mouse clicks and key
presses available in the software application, and therefore the raw number of
interactions in these traces can be large --- containing millions of messages
from different users. Second, for complex software applications, there are
often multiple reasonable interaction paths to accomplish a specific high-level
task while interaction traces that lead to different tasks can share shorter
but common interaction paths.  To address these two challenges of scale and of
uncertainty in interpreting interaction traces, we posit that probabilistic
dimension reduction techniques that can extract frequent patterns from the
low-level interaction data are the right choice to analyze interaction traces.

Topic models are such a dimensionality reduction technique with the capacity to
discover complex latent thematic structures. Typically applied to large textual
document collections, such models can naturally capture the uncertainty in
software interaction data using probabilistic assumptions; however, in cases
where the interaction traces are particularly complex, e.g., in complex software
applications such as IDEs or CAD tools, applying typical topic models may still
result in a large topic space that is difficult to interpret. The special class
of hierarchical topic models encodes a tree of related topics, enabling further
reduction in complexity and dimensionality of the original interaction data and
improving the interpretability of the model. We apply a hierarchical topic
modeling technique, called the Nested Hierarchical Dirichlet Process
(NHDP)~\cite{6802355} to combine interaction traces and stack traces gathered
from a complex software application into a single, compact representation. The
NHDP discovers a hierarchical structure of usage events that has the following
characteristics:

\begin{itemize}
\item provides an interpretable summary of the user interactions that commonly co-occur with specific stack traces;
\item allows for differentiating the strength of the relationship between specific interaction trace messages and a stack trace; and
\item enables locating specific interactions that have co-occurred with numerous runtime errors.
\end{itemize}

\noindent In addition, as a Bayesian non-parametric modeling technique, NHDP
has an additional advantage. It allows the model to grow structurally as it
observes more data. Specifically, instead of imposing a fixed set of topics or
hypotheses about the relationship of the topics, the model grows its hierarchy
to fit the data, i.e., to ``let the data
speak''~\cite{Blei:2010:NCR:1667053.1667056}. This is beneficial in modeling the
datasets of interest since users' interaction with software
changes as the software does, e.g., by adding new features or
removing (or introducing) new bugs. 

%If an interaction (a command or an event)  is commonly associated with a stack
% trace, then the relationship can be further explored to diagnose the problem.

The main contributions of this paper are as follows:
\begin{itemize}
    \item We apply a hierarchical topic model to a large collection of
      interaction and stack trace data produced by ABB RobotStudio, a popular
      robot programming platform developed at ABB Inc, and examine how
      effective it extracts latent thematic structures of the dataset and how
      well the structure depicts a context for exceptions occurring
      during the production use of RobotStudio. 

    \item We are first to propose the idea of grouping users' IDE interaction traces with
        stack traces hierarchically and probabilistically into ``clusters''.
        These ``clusters'' provide user interaction contexts of stack traces.  Since a
        stack trace may be the result of multiple different interaction contexts, this
        approach associates a stack trace with its contexts probabilistically. 

\end{itemize}

We organize the remainder of this paper as follows.
Section~\ref{sec:background} introduces the types of interaction and stack
trace data we use and how we prepare these data sources for topic modeling. We
describe the hierarchical topic modeling technique and its application to
software interaction and crash data in Section~\ref{sec:nhdp}. We apply the
modeling technique to the large RobotStudio dataset and provide an evaluation
in Section~\ref{sec:eval}. Our work is not without threats to its validity, which we discuss in
Section~\ref{sec:threat}. In
Section~\ref{sec:related-work}, we describe relevant related research and
conclude this paper in Section~\ref{sec:end}.

%% file: sec_bg.tex
\section{Background}\label{sec:background}

\begin{figure*}[ht]
   \centering
   \includegraphics[width=\textwidth]{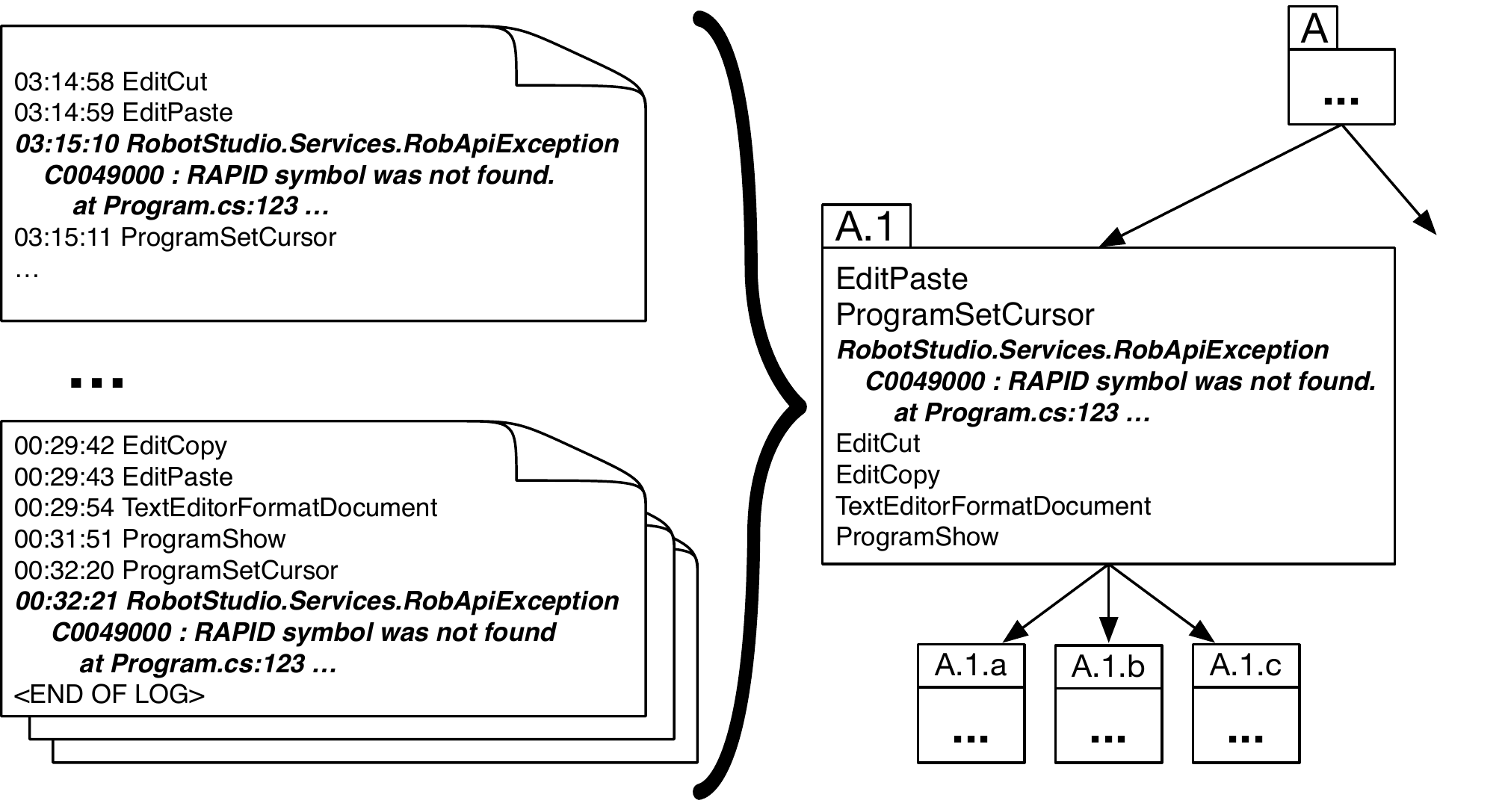}
   \caption{The left half of the figure shows interaction traces with
   embedded stack traces. Note that in the shown interaction traces, the embedded stack
   traces are identical. In this paper, we construct a model that yields a context of the stack
   trace, like the one described in Section~\ref{sec:nhdp}, on the right half of
   the figure.  The model aggregates a collection of interaction traces coupled
   with stack traces into a hierarchy of topics (or contexts). Each topic
   expresses a set of interaction messages with different probabilities,
   depicted via text size in this figure. Note that on the left half of the
   figure each message in the traces has a time stamp depicting the sequence
   their appear while on the right half there are no time stamps. This is because when we apply
   a topic model to the traces, we capture the co-occurring relationship of
   commands and events, such as, the co-occurrence relationship of
   \code{EditCopy} and \code{ProgramSetCursor} but discard the timing
   of the commands and events.}

   \label{fig:data_hists}
\end{figure*}

%basec features of interaction data
Interaction data gathered from complex software applications, such as
IDEs\footnote{The Eclipse UDC dataset is a well known source of this type of
data in the software engineering community. Available at:
\url{http://archive.eclipse.org/projects/usagedata/}}, typically consists of a
large vocabulary of messages, ordered in a time series. The data is typically
collected exhaustively, in order to capture user actions in an interpretable,
logical sequence. As users complete certain actions much more often than others,
the occurrence of interaction messages follows a skewed distribution where some
messages appear often, while most occur infrequently. Some of the messages are
direct results of user actions (i.e., commands), while the others may reflect
the state of the application (i.e., events), such as the completion of a
background task like a project build. Consider the below snippet of an
interaction trace, gathered in Visual Studio, part of the Blaze
dataset~\cite{Snipes:2014:EGD:2591062.2591171,Damevski_Mining_2017}

\begin{center}
\begin{minipage}[t]{0.8\textwidth}
{\small
\begin{verbatim}
2014-02-06 17:12:12  Debug.Start
2014-02-06 17:14:14  Build.BuildBegin
2014-02-06 17:14:16  Build.BuildDone
2014-02-06 17:14:50  View.OnChangeCaretLine
2014-02-06 17:14:50  Debug.Debug Break Mode
2014-02-06 17:15:02  Debug.EnableBreakpoint
2014-02-06 17:15:06  Debug.EnableBreakpoint
2014-02-06 17:15:10  Debug.Start
2014-02-06 17:15:10  Debug.Debug Run Mode
\end{verbatim}
}
\end{minipage}
\end{center}
%goal of extracting high-level behavior from interaction data

The developer that generated the above interaction trace snippet is starting the
interactive debugger, observed by the \code{Debug.Start} command. This triggers
an automatic build in Visual Studio, shown by the \code{Build.BuildBegin} and
\code{Build.BuildDone}, the exact same log messages that appear when the user
explicitly requests the build to begin. After the debugger stops at a
breakpoint, \code{Debug.Debug Break Mode}, this developer enables two
previously disabled breakpoints (e.g., \code{Debug.EnableBreakpoint}) and
restarts (or resumes) debugging (such as, \code{Debug.Start} and
\code{Debug.Debug Run Mode}).

We leverage a probabilistic approach where we model each extracted high-level
behavior as a probability distribution of interaction messages. This type of
model is able to capture the noisy nature of interaction
data~\cite{soh_noises_2015}, which stems from the fact that 1) numerous paths
that represent a specific high-level behavior exist (e.g., using
\code{ToggleBreakpoint} versus \code{EnableBreakpoint} has the same effect) and
2) unrelated events may be in the midst of a set of interactions (e.g.,
\code{Debug.BuildDone} can occur at intervals beginning at
\code{Debug.BuildStart} and interspersed with other messages).

One particular application domain where probabilistic models have been
effective for extracting high-level contexts, or topics, is natural language
processing. In natural language texts words are the most basic unit of the
discrete data and documents can be sets of words (i.e., a ``bag of words''
assumption). We can draw an analogy from the characteristics of interaction
traces to natural language texts, i.e., interaction traces exhibit naming
relations such as synonymy and polysemy similar to those in natural language
texts. A trace often contains multiple different messages that share meaning in
a specific behavioral context, e.g., both the \code{ToggleBreakpoint} and
\code{EnableBreakpoint} events have the same meaning in the same context. This
is similar to the notion of synonymy in natural languages, where different words
can have the same meaning in a given context.  Similarly, IDE commands carry a
different meaning depending on the task that the developer is performing, e.g.,
an error in building the project after pulling code from the repository has a
different meaning than encountering a build error after editing the code base.
This characteristic is akin to polysemy in natural language, where one word can
have different meanings based on its context.

Figure~\ref{fig:data_hists} shows an example of two IDE traces containing both
interactions and stack traces from the ABB RobotStudio IDE.\@ Both of these
traces correspond to user writing a program using a programming language called
RAPID into this environment's editor, and performing common actions like
cutting-and-pasting and cursor movement (i.e., \code{EditCut},
\code{EditPaste}, and \code{ProgramSetCursor}). In both trace excerpts the
users encountered the identical exception, \code{RobApiException [...] RAPID
symbol was not found}, as identified by its type and message. While
corresponding to the same high-level user behavior, the sequence and
constituent messages occurring in the two interaction traces are slightly
different. The modeling approach described here is able to capture the
common interaction context of \code{RobApiException}, forming high-level user
behaviors that we represent as a probabilistic distribution of interaction
messages and stack traces, shown in the right part of Figure~\ref{fig:data_hists}.  The model is
able to overcome the slightly different composition and order in the two
interaction traces, extracting their commonalities, and can help
better characterize and understand the context of the shown exception's stack
trace.

The above motivates us to seek an algorithm to find not only useful sets of
patterns of user behaviors, and learn to organize the these patterns according
to a hierarchy in which more generic or abstract patterns near the root of the
hierarchy and more concrete patterns are near the leaves. This hierarchy would
allow us to  explore stack traces and associated user interactions from the generic to the specific, 
in a way no different from what we do in our daily lives,
i.e. when we go to a grocery store, we begin with a particular section, and then down to a specific aisle, 
finally locating a particular product.

%This understanding can
%have many different applications, such as, to prioritize bug fixes, or to
%cluster bugs, or to reuse bug fix knowledge. We apply NHDP to log traces. NHDP
%is a probabilistic topic models that discover hierarchical structures of
%topics, which provides a contextual information of stack traces.  Blei et al.
%argue that ``if interpretability is the goal, then there are strong reasons to
%prefer'' a hierarchical topic model, such as, hLDA  to non-hierarchical model,
%such as, LDA~\cite{Blei:2010:NCR:1667053.1667056}. NHDP is an extension of
%hLDA. The extension is that NHDP allows each document to exhibit multiple paths
%along the topic hierarchical, a tree. As such, NHDP shows superior modeling
%technique and has potential to achieve high predictive performance and
%interpretability than many known topic models.

\subsection{Topic Models for Interaction Data}
% \item Preliminaries (documents from interaction traces, topic modeling)
Given a collection of natural language documents, topic modeling allows one to
discover latent thematic structures in the document collection (commonly called
a corpus)~\cite{blei2003latent}. A document in the corpus is an unordered set
of words (i.e., ``a bag of words'').  The vocabulary of the corpus, denoted as
$\set{V}$, consists of the $\cardinality{\set{V}}$ unique words in the corpus.
A topic is a discrete probability distribution over the vocabulary words. A collection of
topics describe the extracted thematic structures in the corpus.  For instance, given the vocabulary of a corpus,
denoted as $\set{V} = \{m_1,m_2,\ldots,m_n\}$, a topic is a discrete
probability distribution represented by its probability mass function, $P(m =
m_i) = P_{m_i}$, where $0 \le P_{m_i} \le 1$, $\sum_{i=1}^{|\set{V}|} P_{m_i} =
1$.  Topic models provide means to express the thematic structures in a document and
a document collection, i.e., using topics and the relationship among the topics. For
instance, in Latent Dirichlet allocation (LDA)~\cite{blei2003latent}, a
popular flat (non-hierarchical) topic model, the thematic structures in the document
collection includes the proportions of each topic exhibited in the collection or
in a specific document in the collection.

%%
% HC: Need examples
%%
Topic models are readily applied to other types of data because the models do
not rely on any natural language specific information or assumptions, e.g.,
a language grammar. Examples of data types other than textual data for which
topic modeling has found success include image collections, genetic
information, and social network
data~\cite{4408965,Pritchard945,Wang:2012:TEO:2339530.2339552}.

In this paper, we apply topic models to interaction traces with embedded stack traces.  
We begin by dividing an interaction trace into segments (or windows).  First, we treat each segment as a
``document'' and each command, event, and stack trace as a word. Furthermore, when we examine a small segment of an
interaction trace, we find that a segment consists of usually highly regular and repetitive patterns. This is
likely the result of the following observation of user behavior. Within a small period of time, a
user is likely focusing on a specific task and interacting with a small subset
of the development environment, resulting in segments with a small number of interaction
messages.  In addition, interaction traces exhibit two naming relations, namely
synonymy and polysemy that also exist in natural texts.  The former refers to
that a user can use a command to complete multiple types of tasks, and the
later that the user can accomplish a task via different types of
commands~\cite{Damevski_Mining_2017}.  We posit that these relationships between the interaction types within
small units of IDE usage time mimics the ``naturalness'' of text~\cite{hindle2012naturalness}, which suggests that models used for
analyzing natural language text can be applied to IDE interaction data.  In this paper, interaction trace messages are the words,
segments of interactions messages are the documents, and all of the observed
segments are the corpus of the study. Note that we use the term ``{\em
window}'' to represent a segment as we use the moving window method described
below to divide an interaction trace into segments.

Interaction traces consist of frequently occurring low-level messages
corresponding to 1) user actions and commands (e.g., copying text into the
clipboard, pasting text from the clipboard, building the project); and 2)
events that occur asynchronously (e.g., completion of the build, stopping at a
breakpoint when debugging). The sequential order between the messages is only
relevant to some behaviors, but not to others. For instance, the event
indicating the completion of the build may be important to the next set of
actions the developer performs, or it may be occurring in the background
without import.

In our model, following the ``bag of words'' assumption, we use a tight moving window of interaction messages generated by
an individual developer, but ignore the message order within the window. This is
a reasonable modeling assumption that captures sequential order but resilient to
small permutations in message order within the window. In addition, developer
interaction traces often contain large time gaps, stemming from breaks in the
individual developer's work. To take account of these we force window breaks
when the time between two consecutive messages exceeds a predefined threshold.
An interaction window is a sequence of $N$ messages denoted as $\vect{m} =
(m_1,m_2,\ldots,m_N)$ where $m_N$ is the $N$-th message in the sequence. A
corpus is a set of $M$ windows, denoted as $\set{D} =
\{\vect{m_1},\vect{m_2},\ldots,\vect{m_M}\} = \vect{m}_{1:M}$ where $M =
|\set{D}|$. 

Software exceptions and stack traces, reporting a software fault, which may or
may not be fatal and result in the software to crash, commonly contain a time
stamp and some type of user/machine identifier that tie them to interactions
from the same user. We use a dataset that interleaves the interactions with the
stack traces. We use a window-based modeling technique, as such, minor timing
issues in relating interaction and software crash data become unimportant, as
long as we tie the stack trace of the crash with the relevant window of
interaction messages. Assuming this reasonable assumption holds, we treat the
stack trace as just another message in the interaction log, i.e., the
``vocabulary'' becomes $\set{V} = \{m_1,m_2,\ldots,m_n,s_1,s_2,\ldots,s_p\}$,
where $m$ is an interaction message and $s$ is a stack trace.  Following the
``bag of words'' assumption, we represent document $\vect{m}$ to the
term-frequency form, i.e., $\vect{m}_{tf} = (f_{m_1}, f_{m_2}, \ldots, f_{m_n},
f_{s_1}, f_{s_2}, \ldots, f_{s_p})$ where $f_w$ is the frequency of word $w$,
either an interaction message or a stack trace in
vocabulary $\set{V}$.

%In addition, interactions traces exhibit some properties similar to natural
%languages, such as synonymy and polysemy. In a natural language
%document, many different words can have the same or similar meaning.
%Similarly, when a user interacts with a complex software application (e.g. an IDE),
%she can accomplish the same
%task using different combinations of commands and interactions. For instance,
%she can start a debug session via shortcuts keys or via a context menu.
%An additional observation is that in a natural language document the same word can have different
%meaning depending on the context.  Similarly, when a developer is interacting
%with the IDE, she may use the same command to achieve many different purposes.
%These properties in topic modeling are revealed via co-occurring words, and the
%co-occurrence relationship is reflected in the topic structure.

%% file: sec_nhdp.tex
\section{Hierarchical Topic Modeling for Interaction Data}\label{sec:nhdp}

The scale of IDE interaction traces collected from the field can pose a
challenge to analysis. The size of the traces can grow quickly and become 
large, for instance, the Eclipse Foundation Filtered UDC Data set consists of on
the order of ~$10^7$ messages a day. Our approach is to divide the traces into
message windows. To accomplish this, we first divide the traces into active
sessions, using a prolonged time gap between messages as a delimiter, and
further divide each session into one or more windows, each of which is a
sequence of a fixed number of messages. Stack traces appear in the
interaction log from time to time. We treat them as ordinary messages in the
windows in the model. In the remainder of the paper, to be consistent with prior
literature on topic models, we sometimes refer to a message window as a {\em
document}, and messages within that window as {\em words}.

Our windowing approach bears similarity to the data processing method
commonly used for streaming text corpora, such as, transcripts of automatic
speech recognized streaming audio, transcripts of closed captioning, and feeds
from the news wire~\cite{Blei:2001:TSA:383952.384021}. Among these kinds of
datasets, no explicit document breaks exist. A common approach is to divide the
text into ``documents'' of a fixed length, as we have.

%Since implicit
%representation of document breaks exists, a few approaches have been proposed
%to combine adjacent documents between which there are no implicit breaks, as a post-processing
%step to the window creation. These
%methods typically rely on a human-labeled corpus of
%document breaks to train a model, which is subsequently used to detect document
%breaks~\cite{Blei:2001:TSA:383952.384021,purver2011topic}. This approach is challenging
%due to the large amount of data we aim to analyze, necessitating a sufficiently large
%training set.

%\subsection{Nested Hierarchical Dirichlet Process}\label{subsec:nhdp}

Most topic models, such as LDA, are flat topic models, in which the topics are
independent and there is no structural relationship among the discovered
topics. There are two challenges facing flat topic models. First, it is
difficult or at least computationally expensive to discover the number of
topics that we should model in a document collection.  Second, since there is
only a rudimentary relationship among topics, the meaning of the topics is
difficult to interpret, in particular, when multiple topics may look alike
based on their probability distributions.

%
% HC: the method uses Chinese Restaurant Process. However, if the messages
% in the traces follow the Power law, the Pitman-Yor process may be
% more appropriate.
%
We use a hierarchical topic model based on the Nested Hierarchical Dirichlet
Process (NHDP), which, compared with a flat topic model, arranges the topics in
a tree where more generic topics appear on upper levels of the tree while more
specific topics appear at lower levels.  We can achieve two objectives via a
hierarchical topic model. The number of topics for a model can be easily
expressed in the hierarchy, much like the hierarchical clustering algorithm
where we can determine the number of clusters by increasing gradually the depth
and the branches of the tree of clusters.  In addition, the hierarchical
structure of the topics, i.e., more generic topics appearing on upper levels
of the tree and more specific topics on lower levels can lead to improved human
interpretability. As argued in~\cite{Blei:2010:NCR:1667053.1667056}, ``if
interpretability is the goal, then there are strong reasons to prefer'' a
hierarchical topic model, such as, NHDP over a flat topic model, such as, LDA.\@

A number of hierarchical topic models exist in the literature. We choose the
Nested Hierarchical Dirichlet Process
(NHDP)~\cite{Blei:2010:NCR:1667053.1667056} as it possesses some advantages
over other popular hierarchical models, such as the Hierarchical Latent
Dirichlet Allocation (HLDA)~\cite{Blei:2010:NCR:1667053.1667056}.  Different
from these models, NHDP results in a more compact hierarchy of topics (less
branching) and produces less repetitive topics as it allows a document to
sample topics from a subtree that is not limited to a path from the root of the
tree. For the IDE interaction traces of our interest, NHDP is a right modeling
tool because a stack trace can occur at different interaction contexts and we
can capture this variability effectively at higher (more general) levels of its
hierarchy and differentiate the contexts at lower (more specific) level of the
hierarchy.

\begin{figure}[ht]
\begin{center}
\begin{tikzpicture}
    \tikzstyle{every node}=[font=\footnotesize]

    \tikzstyle{hyperparameter} = [circle, minimum width=0.275in, draw=black,
                        label={below:#1}]

    \tikzstyle{latentnode} = [circle, minimum width=0.275in, draw=black,
                        label={below:#1}]

    \tikzstyle{datanode} = [circle, minimum width=0.275in, draw=black,
                        fill=gray!20, label=below:#1]

    \tikzstyle{plate} = [rectangle,
                        text centered, draw=black,
                        inner sep=0pt,
                        label={[xshift=-16pt,yshift=13pt]south east:#1}]

    \tikzstyle{connect} = [arrows=->, black]

    \node(word)[datanode,label={center:$w_{d,n}$}]{};
    \node(topic)[latentnode,label={center:$\theta_k$}
        ,right=0.35in of word]{};
    \node(topicplate)[plate
        ,label={[xshift=-14pt,yshift=10pt]south east:$\infty$}
        ,fit=(topic)
        ,inner sep=6pt]{};
    \node(eta)[hyperparameter ,label={center:$\eta$}
        ,right=0.30in of topic]{};
    \node(topicindicator)[latentnode,label={center:$c_{d,n}$}
        ,left=0.15in of word]{};
    \node(doctopictree)[latentnode,label={center:$\mathcal{T}_d$}
        ,above left=0.10in and 0.20in of topicindicator]{};
    \node(levelswitch)[latentnode,label={center:$U_d$}
        ,below left=0.10in and 0.20in of topicindicator]{};
    \node(globaltopictree)[latentnode,label={center:$\mathcal{T}$}
        ,left=0.27in of doctopictree]{};
    \node(doctopictreeconcentration)[hyperparameter,label={center:$\beta$}
        ,above=0.30in of doctopictree]{};
    \node(dpbasemeasure)[hyperparameter,label={center:$\mathbf{H}$}
        ,above=0.30in of globaltopictree]{};
    \node(dpconcentration)[hyperparameter,label={center:$\alpha$}
        ,left=0.30in of globaltopictree]{};
    \node(levelswitchalpha)[hyperparameter,label={center:$\gamma_1$}
        ,above left=0.05in and 0.35in of levelswitch]{};
    \node(levelswitchbeta)[hyperparameter,label={center:$\gamma_2$}
        ,below left=0.05in and 0.35in of levelswitch]{};
    \node(docplate)[plate
        ,label={[xshift=-16pt,yshift=11.5pt]south east:$N_d$}
        ,fit=(topicindicator) (word)
        ,inner sep=8pt]{};
    \node(corpusplate)[plate
        ,label={[xshift=-14pt,yshift=10pt]south east:$M$}
        ,fit=(doctopictree) (levelswitch) (topicindicator) (word) (docplate)
        ,inner sep=8pt]{};

    \draw[connect](eta) to (topic);
    \draw[connect](topic) to (word);
    \draw[connect](topic) to (word);
    \draw[connect](topicindicator) to (word);
    \draw[connect](doctopictree) to (topicindicator);
    \draw[connect](levelswitch) to (topicindicator);
    \draw[connect](globaltopictree) to (doctopictree);
    \draw[connect](dpbasemeasure) to (globaltopictree);
    \draw[connect](dpconcentration) to (globaltopictree);
    \draw[connect](levelswitchalpha) to (levelswitch);
    \draw[connect](levelswitchbeta) to (levelswitch);
    \draw[connect](doctopictreeconcentration) to (doctopictree);

\end{tikzpicture}
\end{center}
    \caption{The probabilistic graphical model of NHDP.\@ The model is a
    Bayesian network, represented as a directed graph.  There are $3$ plates in
    the graph, the topic plate that represents potentially infinite number of
    topics, the document plate for a document and the corpus plate.  We denote
    the number of words in document $d$ as $N_d$. The corpus has $M$ documents.
    The $n$-th word in document $d$, $w_{d,n}$ is the only observable variable
    in the model.  For $n$-th word in document $d$, we draw a topic indicator
    based on the topic tree $\mathcal{T}_d$ and the switching probabilities
    $U_d$, where we draw $\mathcal{T}_d$ from global topic tree $\mathcal{T}$
    and draw $U_d$ from a Beta distribution with two hyperparameters $\gamma_1$
    and $\gamma_2$.  } 
    \label{fig:nhdp:pgp}

\end{figure}
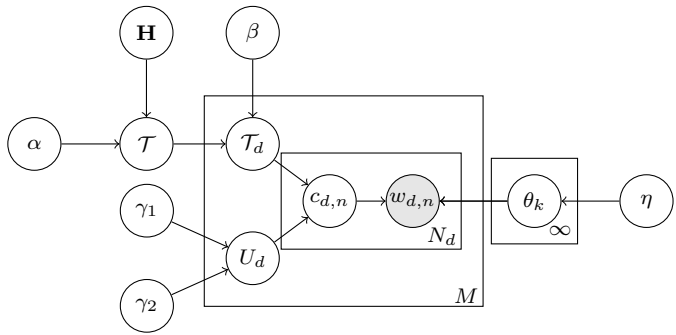

To understand how we may apply the NHDP topic model to analyze software
interaction traces, we illustrate the model in Figure~\ref{fig:nhdp:pgp} as a
directed graph, i.e., a Bayesian network.  Since NHDP is a Bayesian model, it
starts with a {\em prior}.  In effect, the name of the NHDP topic model comes
from that of its prior, i.e., the nested hierarchical Dirichlet process.  The
prior expresses the assumption that the thematic structure of the topics is in
a tree-like structure and the assumption that a topic can have branches
corresponding to more specific topics at lower level in the tree.  We
specify or tune these assumptions by giving a number of parameters of
the prior as inputs to the model, commonly
referred to as the hyperparameters of the model.  We provide an overview of
these hyperparameters and their relationship with other variables in the graph
in Figure~\ref{fig:nhdp:pgp}.

In NHDP, we consider words in documents to follow Multinomial distributions,
given a topic. Dirichlet distributions are a commonly used prior for
multinomial distributions. It follows that we draw topics, a set of multinomial
distributions over words from given Dirichlet distributions.  As shown in
Figure~\ref{fig:nhdp:pgp}, given a hyperparameter $\eta$ as the parameter for a
Dirichlet distribution, we draw potentially infinite number of topics, denoted
as $\theta_k$, $k =\{1, 2, \ldots\}$ in Figure~\ref{fig:nhdp:pgp}. Since we
choose a symmetric Dirichlet distribution for generating topic distributions
for this work, hyperparameter $\eta$ is a positive scalar, and represents the
concentration parameter of the Dirichlet distribution $Dir(\eta)$. The smaller
$\eta$ is, more concentrated on fewer words we believe a topic to be.

A topic corresponds to a node in global topic tree $\mathcal{T}$.  We can
either draw a global topic tree $\mathcal{T}$ from a nested Chinese Restaurant
Process as illustrated in~\cite{Blei:2010:NCR:1667053.1667056} or construct it
directly using a nested Stick Breaking Process as shown in~\cite{6802355}. Both
of these two methods yield an infinite set of Dirichlet processes, each
corresponding to a node in the tree.  A Dirichlet process, an infinitely
decimated Dirichlet distribution, allows us to branch from a topic node to an
infinite number of child topic nodes, which constitutes the mechanism to build
the topic tree.  A Dirichlet process is a distribution from which a draw is
also a probability distribution.  We denote drawing a probability distribution
$G$ from a Dirichlet process as $G \distas{} DP(\alpha\mathbf{H})$ where
concentration parameter $\alpha$ and base measure $\mathbf{H}$ are two
hyperparameters as shown in Figure~\ref{fig:nhdp:pgp}.  The probability
distributions drawn from the Dirichlet process provide a parameter to associate
a node in the topic tree to its corresponding topic ($\theta_k$).  The
concentration parameter $\alpha$, where $\alpha > 0$ represents our belief on
how we should branch a topic node to topic nodes on a lower level. The
greater the $\alpha$, the more branches we should expect when given a corpus.

When examining the relationship of the topics, we know that the topics depend
on the manner that we derive document trees in the model. A document tree
$\mathcal{T}_d$ is a copy of the global topic tree of the given corpus with the
same topics on the nodes but with different branching probabilities.  As
discussed above, an important characteristic of NHDP is its prior, the nested
hierarchical Dirichlet process that leads to the mechanism by which we branch a
topic node to a lower level. Each node in the global tree has a corresponding
Dirichlet process. Let's denote the Dirichlet process at a node $n$ in the
global tree $\mathcal{T}$ as $G_{\mathcal{T}_n}$, the corresponding node in the
topologically identical document topic tree for document $d$ has a Dirichlet
process $G_{d} \distas{} DP(\beta G_{\mathcal{T}_n})$, where the concentration
parameter $\beta$ controls our belief on how a document branches in the
corresponding document tree, i.e., hyperparameter $\beta$ controls how the
branching probability mass is distributed among branches.  The higher the
$\beta$, the less concentrated the branching probability mass is, and in effect,
the more branches we should expect from a corpus. For instance, if we expect a
document in the corpus should branch to a small number of topics in next
level, all the while we expect these topics to be different among different
documents, we should begin with a large $\alpha$ and a small $\beta$ because we
expect {\em effectively} a large global tree, but small document trees.

Furthermore, each word in a document has a topic since we sample words from a
topic, i.e., a discrete probability distribution over words. We conveniently refer a topic
by using its index. Denote the index of the $n$-th word's topic in document $d$
as $c_{d,n}$ as shown in Figure~\ref{fig:nhdp:pgp}. We determine the topic for
the word from a two-step approach. First, we choose a path from the root in the
document tree $\mathcal{T}_d$ based on the tree's branching probabilities.
Next, we select a topic along the path for the word based on a probability
distribution --- starting from the root along the path, we draw $U_d$ from Beta
distribution $Beta(\gamma_1, \gamma_2)$, and $U_d$ is the probability that we
remain on the node, and $1 - U_d$ is the probability that we switch to next
node along the path. The two parameters control the expected range of the level
switching probabilities. The Beta distribution here is commonly used to express
a probability distribution of probabilities.

These hyperparameters have an impact on the learned NHDP model and inference of
new documents. In Section~\ref{sec:eval}, we evaluate how sensitive the
learned NHDP model is to the hyperparameters. An insensitive model has stronger
ability to correct inaccurate hyperparameter priors by learning what the data implies.

\subsection{Learning the NHDP Model}

To learn a NHDP model from a document corpus, we adopt the stochastic inference
algorithm in~\cite{6802355}. The algorithm has the following steps:

\begin{enumerate}
    \item \label{alg:infer:vocab} Scan the documents from the training corpus,
        and extract words to form a vocabulary of the training corpus. In
        this step, the vocabulary consists of IDE messages and stack traces. 
        We treat a stack trace as a single word. Denote the vocabulary 
        as $\set{V}$ that consists of $\cardinality{\set{V}}$ unique words.

    \item \label{alg:infer:tf} Index words in the vocabulary from $0$ to
      $\cardinality{\set{V}} - 1$, and convert each document to a {\em
      term-frequency} vector where the value at position $i$ is the frequency
      of the word indexed by $i$ in the document.

    \item \label{alg:infer:init} Randomly select a small subset of documents
        from the training corpus, denote the set of documents as $D_{I}$. The
        random selection of documents will not stop until 
        any word in the vocabulary appears at
        least once in the selected documents.

    \item \label{alg:infer:cluster} Repeatedly run the 
        $K$-means clustering algorithm against $D_{I}$ to build a
        tree of clusters.

    \item \label{alg:infer:tree} Initialize a NHDP tree for $D_{I}$, call the
        initial NHDP topic tree as $\mathcal{T}_{I}$, and let
        $\mathcal{T}_{R} = \mathcal{T}_{I}$.

    \item \label{alg:infer:step:init} Randomly select a subset of documents
        from the training corpus, denote the set of documents as $D_{R}$.

    \item \label{alg:infer:step:infer} Make adjustment to $T_{R}$ based on an
        inference algorithm against $D_{R}$. The result is a topic tree
        $\mathcal{T}$.

    \item \label{alg:infer:step:loop} Repeat steps~\ref{alg:infer:step:init}
        and~\ref{alg:infer:step:infer} until $\mathcal{T}$ converges.

\end{enumerate}

From steps~\ref{alg:infer:init} to~\ref{alg:infer:tree}, we provide the maximum
height and the maximum number of nodes at each level of tree $D_{I}$.  The
maximum height and number of nodes at each level should be greater than the
final tree. Following the assumption that words are interchangeable, we
convert a document to the term-frequency form, i.e., 
a vector where each element is the frequency of the
corresponding word appearing in the document. In Step 4, we use the K-means
clustering algorithm to divide the documents into a number of clusters, and for
each cluster, we estimate a topic distribution.  These clusters and the topic
distributions are the top level nodes in tree $D_{I}$ just beneath the root.
We then repeat the process for each cluster, and each cluster is further
divided into a number of subclusters. For each subcluster we estimate a topic
distribution. This step is for computational efficiency. Given the number of
clusters and the depth of the tree, the $K$-means algorithm builds a large tree
quickly. This tree serves as the initial tree for the NHDP algorithm that
learns the switching probabilities for different levels and the switching
probabilities for different clusters at a level, which effectively shrinks the
tree by learning the switching probabilities.  Note in the above when applying
the K-means algorithms, we adopt the $L_1$ distance, i.e., given two documents
represented as two vectors $\vec{d}_i$ and $\vec{d}_j$, the distance of the two
documents is $d(\vec{d}_i, \vec{d}_j) = \sum_{k=0}^{\cardinality{\set{V}}-1}
|d_{ik} - d_{jk}|$.

Steps~\ref{alg:infer:step:init} to~\ref{alg:infer:step:loop} perform a
randomized batch inference processing.  Agrawal et al.\ demonstrate that topic
modeling can suffer from ``order effects'', i.e., a topic modeling algorithm
yields different topics when we alter the order of the training
data~\cite{agrawal2018wrong}. This randomized batch processing can reduce this
``order effects''  via averaging over different random orders of the
training data set.  Step~\ref{alg:infer:step:infer} requires a specific
inference algorithm.
In~\cite{Blei:2010:NCR:1667053.1667056,doi:10.1198/016214506000000302}, Markov
Chain Monte Carlo algorithms, specifically, Gibbs samplers are used. In this
work, we used the variational inference algorithm in~\cite{6802355}.
Variational inference algorithms are typically shown to scale better to large
data sets than Gibbs samplers do.  Steps~\ref{alg:infer:step:init}
to~\ref{alg:infer:step:loop} can begin with an arbitrary tree, however, it is
much more computationally efficient to initialize the inference algorithm with
a tree that shares statistical traits with the target data.

%% file: sec_eval.tex
\section{Evaluation}\label{sec:eval}

%\begin{enumerate}
%    \item Using held out documents (automatic/machine-based evaluation)
%        \begin{enumerate}
%
%        \item Compute likelihood/perplexity
%
%        \item Computer prediction likelihood/perplexity (document level prediction: split each document in held out documents into 90% + 10%, train the model using training dataset; predict 10% for all held documents, and compute likelihood)
%        \end{enumerate}
%
%    \item Survey meant to understand if useful information can be gained from approach (human-based evaluation)
%\end{enumerate}

For evaluation, we use field interaction traces from ABB RobotStudio, a popular
IDE intended for robotics development that supports both simulation and
physical robot programming using a programming language called RAPID.\@
RobotStudio as an IDE also runs robot application programs developed in the IDE
by users. It is RobotStudio that collects interaction traces other than the
robot applications do.  The RobotStudio interaction trace dataset we used
represents $25,724$ users over a maximum of $3$ months of activity, or a total
of $76,866$ user-work hours. In the interaction traces, there are $7,425$
unique messages, $134$ types of exceptions, $1,975,474$ sessions, and $2,251$
unique stack traces, resulting in $1,978,081$ documents of $50$ messages. Note
that a single exceptions in RobotStudio is often triggered by numerous users of
the IDE, as such, an exception corresponds to many unique stack traces and each
unique stack trace has many copies.  We chose the window size of $50$ messages
based on empirically observing this to result in semantically interesting
windows, which commonly represent a single activity by a
developer~\cite{Damevski_Predicting_2017}.

%% why not use sessions? since the number of sessions is not too much greater
%% than the number of windows?
%% no message (exception removed)
%interaction traces, there are $3,727$ unique messages, and $1,983,317$
%documents of $50$ messages.

The RobotStudio dataset consists of sequences of time-stamped messages, where
each message corresponds to a RobotStudio command (e.g.,
\code{RapidEditorShow}) or an event representing application state (e.g.,
\code{Exception} and \code{StartingVirtualController}). Messages have
additional attributes, such as the component that generates the command or the
event, and the command or event type. RobotStudio records the stack traces
directly into the interaction log, so the two distinct data types considered
here are already combined into one single trace.

The evaluation plan is as follows. First, we conduct a ``held-out'' document
evaluation, i.e., we divide the documents into two sets, training dataset to
learn the model, and held-out dataset to test the model. The purpose of the
held-out document evaluation are two-fold. We want to know whether the training
data set is sufficient to produce a stable model and to assess whether the
parameters used in the learning process is reasonable.  Second, we conduct a
user survey to assess the usefulness of the model in understanding and
debugging software faults.  Figure~\ref{fig:pipeline} illustrates the
overall processing pipeline used for evaluation. 

\input{flowchart}

\subsection{Held-out Document Evaluation}

Unsupervised learning algorithms, like NHDP, are typically more challenging to
evaluate, as there is no ground truth to compare to.  Perplexity and predictive
likelihood are two standard metrics for informational retrieval evaluation that
corresponds to a model's ability to infer an unseen document from the same
corpus. These two are a single metric in two different representations since
perplexity is, in effect, the inverse of the predictive power of the model. The
worse the model is, the more perplexed it is with unseen data, resulting in
greater values for the perplexity metric. Similarly, the better the model is,
the more likely that the model is able to infer the model of an unseen
document. To further explain these two concepts and their relationship and how
we compute them, let us divide the dataset into two subsets, one is a
training dataset that we consider as observed, and the other a held-out
dataset that we consider as unseen. We denote the former as $D_{\mobs}$ and
later as $D_{\mheldout}$. We consider $D_{\mobs}$ has $N_{\mobs}$ documents,
and $D_{\mobs} = \{d_{\mobs,1}, d_{\mobs,2}, \ldots, d_{N_{\mobs}}\}$, and
$D_{\mheldout}$ has $N_\mheldout$ documents, and $D_{\mheldout} =
\{d_{\mheldout,1}, d_{\mheldout,2}, \ldots, d_{\mheldout,N_\mheldout}\}$. Given
that we learn a model $\mathcal{M}$ from the training dataset $D_{\mobs}$, we
define the predictive power of the learned model is the following conditional
probability, i.e., the probability of observing the unseen documents given the
model learned from the observed document,

\begin{align}\label{eqn:pp}
    &P(D_{\mheldout}|D_{\mobs}, \mathcal{M})  \nonumber \\
        &\quad =  P(d_{\mheldout,1}, \nonumber \\
        &\quad\quad\quad
                \ldots d_{\mheldout,N_{\mheldout}}
                    |
                d_{\mobs,1},
               \ldots,
                d_{\mobs,N_\mobs}, \mathcal{M})\nonumber\\
        &\quad =
            \prod_{i=1}^{N_\mheldout} P(d_{\mheldout,i}
            |
                d_{\mobs,1},
                \ldots,
                d_{\mobs,N_\mobs}, \mathcal{M})
\end{align}

\noindent where we assume that held-out documents are independent to one
another.

Since the probability in equation~\eqref{eqn:pp} varies on the size of the
held-out dataset, $N_\mheldout$, the probability is not comparable for
held-out datasets of different sizes.  To make it comparable among held-out
dataset of different sizes, we take a geometric mean of the probability as
follows,

\begin{align}
    &\overline{P}(D_{\mheldout}|D_{\mobs}, \mathcal{M}) \nonumber \\
    &\quad\quad= P(D_{\mheldout}|D_{\mobs},
        \mathcal{M})^{\frac{1}{\sum_{i=1}^{N_\mheldout}|d_{\mheldout,i}|}}
%    &= p(d_{\mheldout,1},
%            \ldots,
%            d_{\mheldout,n_{\mheldout}}
%            |
%            d_{\mobs,1},
%           \ldots,
%            d_{\mobs,n_\mobs}, \mathcal{m})^{\frac{1}{n_\mobs}}
\end{align}
\noindent where $|d_{\mheldout,i}|$ is the sum of all word counts in document
$d_{\mheldout,i}$.

We call $\overline{P}(D_{\mheldout}|D_{\mobs}, \mathcal{M})$ the predictive
likelihood of the model $\mathcal{M}$ on the unseen dataset $D_{\mheldout}$.
We can then define the predictive log likelihood as,

\begin{align}
    &\mathcal{L}(D_{\mheldout}|D_{\mobs}, \mathcal{M}) \nonumber \\
    &\quad = \log \overline{P}(D_{\mheldout}|D_{\mobs}, \mathcal{M}) \nonumber \\
    %&\quad =\log P(D_{\mheldout}|D_{\mobs},
    %    \mathcal{M})^{\frac{1}{ \sum_{i=1}^{N_\mheldout}|d_{\mheldout,i}|}} \nonumber \\
    &\quad =\frac{1}{ \sum_{i=1}^{N_\mheldout}|d_{\mheldout,i}|}
        \log P(D_{\mheldout}|D_{\mobs}, \mathcal{M})
\end{align}

\noindent and define the perplexity as the inverse of the predictive
likelihood,

\begin{align} \label{eqn:perplexity}
    &Perplexity(D_{\mheldout}|D_{\mobs}, \mathcal{M}) \nonumber \\
    &\quad= \frac{1}{\overline{P}(D_{\mheldout}|D_{\mobs}, \mathcal{M})} \nonumber \\
    %&\quad= P(D_{\mheldout}|D_{\mobs}, \mathcal{M})^{-\frac{1}{ \sum_{i=1}^{N_\mheldout}|d_{\mheldout,i}|}} \nonumber \\
    %&\quad= e^{\log P(D_{\mheldout}|D_{\mobs}, \mathcal{M})^{-\frac{1}{ \sum_{i=1}^{N_\mheldout}|d_{\mheldout,i}|}}} \nonumber \\
    %&\quad= e^{-\frac{1}{ \sum_{i=1}^{N_\mheldout}|d_{\mheldout,i}|} \log P(D_{\mheldout}|D_{\mobs}, \mathcal{M})} \nonumber \\
    &\quad= e^{- \mathcal{L}(D_{\mheldout}|D_{\mobs}, \mathcal{M})}
\end{align}
\noindent which establish the correspondence between perplexity and
predictive log likelihood.

In the following, we describe the procedure to compute the perplexity and show
the result. This evaluation method, inspired by earlier work
in~\cite{Wallach:2009:EMT:1553374.1553515,Rosen-Zvi:2004:AMA:1036843.1036902},
is frequently used to evaluate topic models, such as
in~\cite{6802355,NIPS2014_5303}. The procedure below is adopted from~\cite{6802355}.

\begin{figure}[ht]
    \includegraphics[width=0.95\columnwidth]{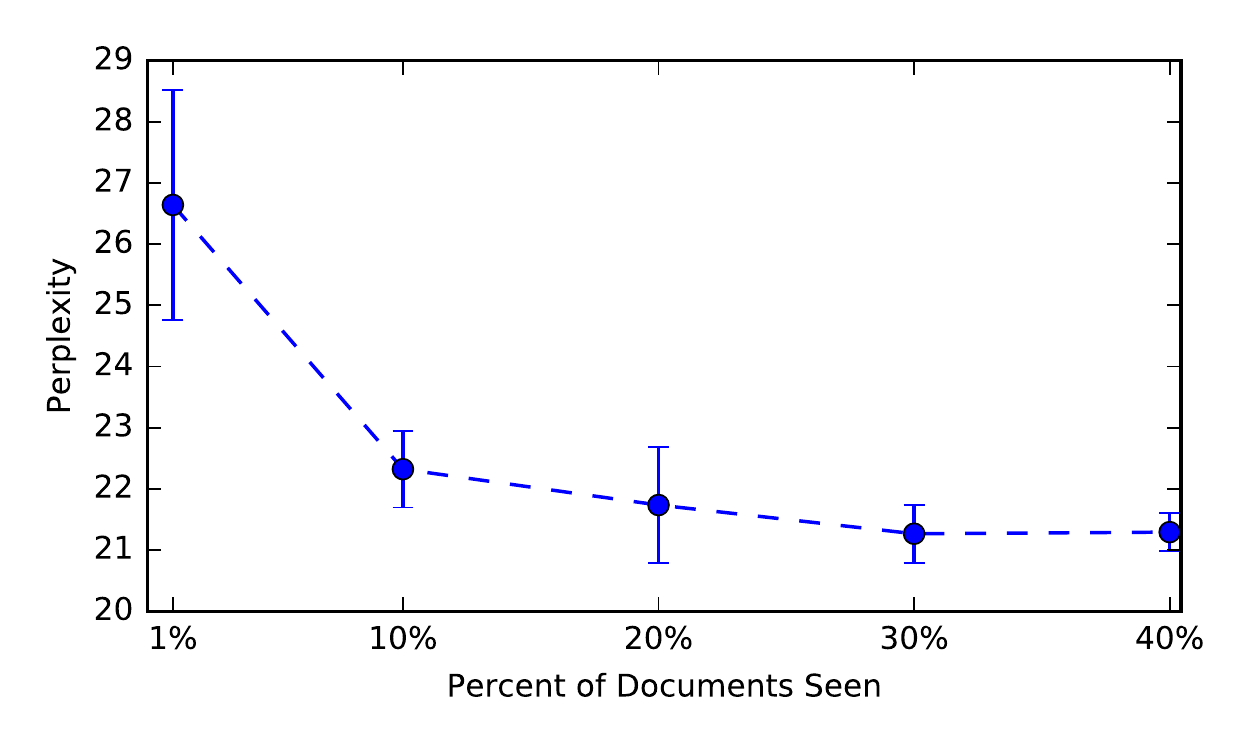}
    \caption{Perplexity versus percent of documents seen. For each
    number of document seen, the standard deviation of the perplexities
    of $10$ runs is also shown. The graph indicates the convergence of
    the training process to a stable model.}
    \label{fig:perplexity}
\end{figure}

\begin{enumerate}
    \item {\bf Form training and testing datasets.} We divide interaction traces
        into a training dataset and a testing dataset based on a reasonable ratio
        $r_{td}$, e.g., $0.9$. To obtain the training dataset, randomly select
        $r_{td} M$ documents from the $M$ documents of interaction traces. The
        remaining  $(1 - r_{td}) M$ documents are in the testing dataset.

    \item {\bf Form observed dataset and held-out dataset.} Select a document
        partition ratio $r_{dp}$, e.g., $0.9$.  For each document $d$ in the
        testing dataset, and the $F_d$ appearances of words in the
        document, partition $F_d$ into two partitions. The first $r_{dp} F_d$ words goes to
        the first partition, and the second $(1 - r_{dp}) F_d$ words the second
        partition. Consider the two partitions as two documents, $d_h$ and
        $d_o$. Then all the $d_h$ form the held-out dataset and all the
        $d_o$ forms all the observed dataset, i.e., we obtain $D_{\mheldout}$
        and $D_{\mobs}$ in equation~\eqref{eqn:perplexity}.

    \item {\bf Train the model.} Use NHDP on the training dataset, i.e., infer
        the global topic tree $\toctree{T}$ using the training dataset.  The
        model is $\mathcal{M}$ in equation~\eqref{eqn:perplexity}.

    \item {\bf Compute perplexity.} Use the definition in
        equation~\eqref{eqn:perplexity}.

\end{enumerate}
Figure~\ref{fig:perplexity} is a result of the perplexity obtained when we
gradually increase the number of documents seen and the use the rest as the
testing data. We take an approach inspired by $N$-fold cross-validation.
For each training dataset size, we randomly select the training dataset from
the collected dataset and then compute the perplexity. We illustrate $10$
computed perplexities at each training dataset size in an $x$-$y$ plot with
error bar in Figure~\ref{fig:perplexity}. The figure shows that both the
perplexity and the variation of the perplexity decreases as training dataset
size increases, indicative of the convergence of the algorithm and a stable model. In particular, when the document seen is at $40\%$ of
documents, we observe a significant drop of perplexity, and the magnitude
of the drop is consistent with those in the topic modeling literature, such
as,~\cite{NIPS2014_5303,6802355,Blei:2010:NCR:1667053.1667056}. This
suggests that the obtained model has converged to a stable state and the model
provides a stable representation of the underlying data. We can now use the
model for the purpose of interpreting the context of software exceptions.

\subsection{Sensitivity Analysis}
As a Bayesian hierarchical model, for NHDP, we infer marginal and conditional
probability distributions from the data for the parameters in the model,
as such, the model does not overfit. As a non-parametric model, we parametrize
the model with infinite number of parameters, as such, the model does not
underfit~\cite[page 101]{gelman2014bayesian}.

One challenge is that we specify the prior of a Bayesian non-parametric model
by giving the values of the hyperparameters of the prior and the values are
sometimes difficult to choose.  We ought to assess the effect of these values.
A common method is via sensitivity analysis.  This is particularly important
for Bayesian hierarchical models~\cite{roos2015sensitivity}.  For sensitivity
analysis, we examine how the hierarchy obtained varies with hyperparameters in
the prior. Their values control the base distribution in the NHDP process, and
the switching probabilities between levels of the tree. For a document, we
draw the topics at a node from a Dirichlet distribution, specifically, draw
them from $Dir(\eta)$, a symmetric Dirichlet distribution controlled by the
concentration parameter $\eta$; however, we need to choose which branch to
visit to draw topics for its children, for which we must know 
hyperparameter $\beta$.  When we generate a document, we decide whether to go
to next level of the tree based on Beta distribution, $Beta(\gamma_1,
\gamma_2)$. We explain the effects of these parameters in
Section~\ref{sec:nhdp}.

\begin{figure}[ht]
    \includegraphics[width=0.95\columnwidth]{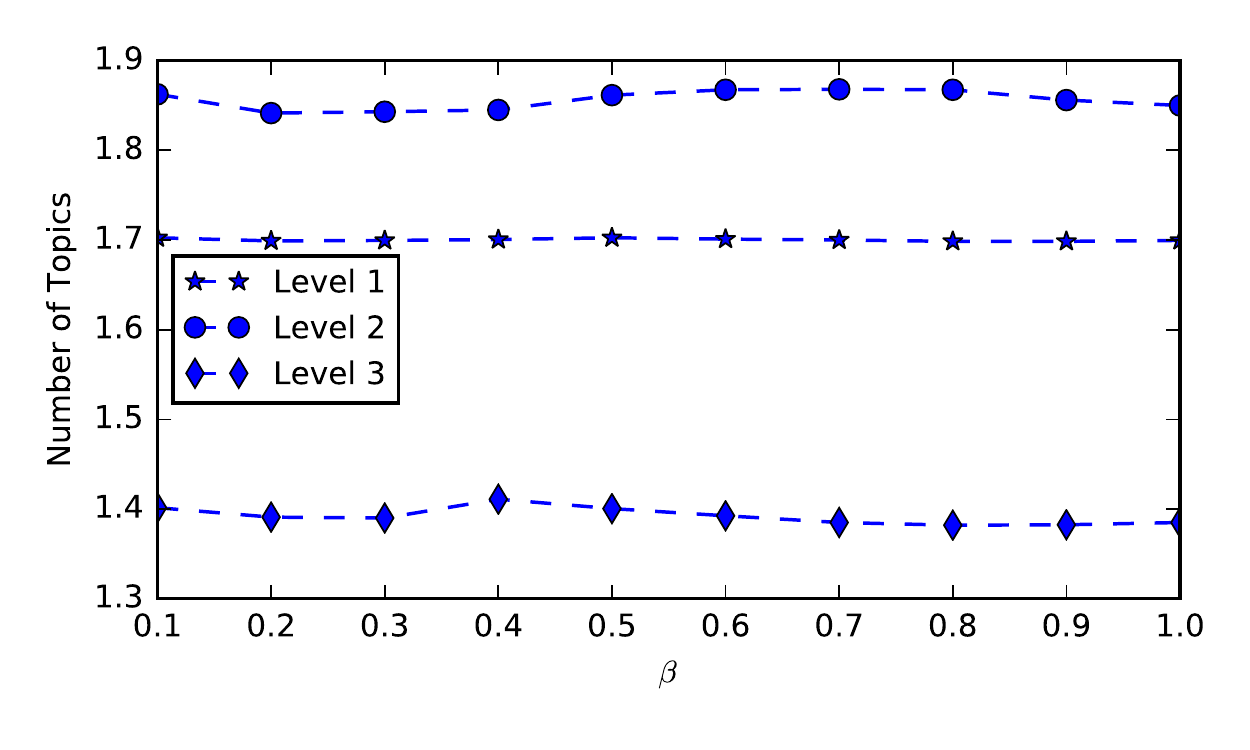}
    \caption{The average number of topics per document at
    tree levels $1$ -- $3$ versus hyperparameter $\beta$. The graph indicates
    the desired characteristic of the model that it is insensitive to the
    hyperparameter $\beta$ of the prior.}
    \label{fig:topicsvsbeta}
\end{figure}

We use a number of statistics to evaluate how sensitive the learned model
is to the hyperparameters. These statistics include the number of topics at each
level of the tree for each document and the number of words at each topic.
Figure~\ref{fig:topicsvsbeta} shows the average number of topics per document
at tree levels 1, 2, and 3 when we increase hyperparameter $\beta$ from $0.1$ to
$1.0$ when we infer the model from a set of $40\%$ of randomly selected
documents. The graph shows that the inferred model is insensitive to
the hyperparameter $\beta$.

\begin{figure}[ht]
    \includegraphics[width=0.95\columnwidth]{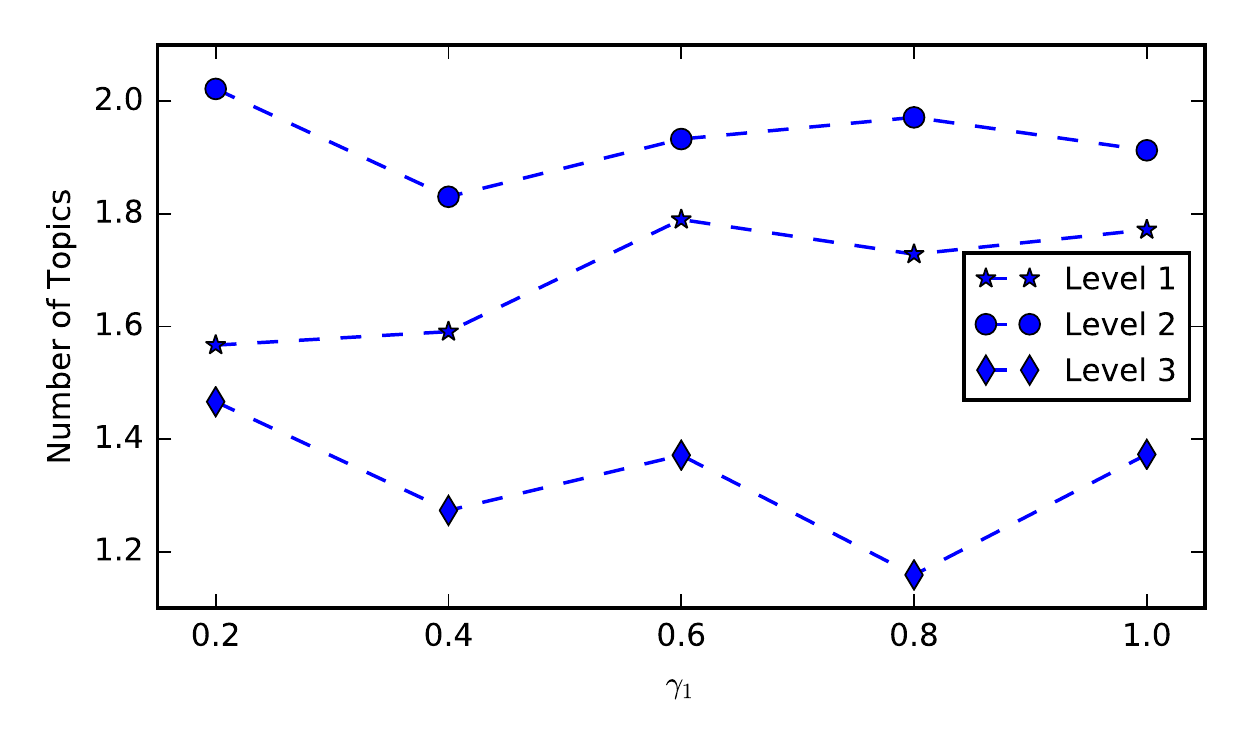}
    \caption{The average number of topics per document at tree levels $1$ --
    $3$ versus $\gamma_1$. When we vary $\gamma_1$, we hold $\gamma_1 +
    \gamma_2 = 2$. When $\gamma_1 = \gamma_2 = 1$, the Beta distribution
    becomes a uniform distribution in $[0, 1]$. As $\gamma_1$ increases,
    we become less likely to draw smaller probabilities from the Beta
    distribution, which results in words more likely
    to stay on the current level of the tree.}
    \label{fig:topicsvsgamma}
\end{figure}

Figure~\ref{fig:topicsvsgamma} shows the average number of topics per document
at tree levels 1, 2, and 3 when we increase hyperparameter $\gamma_1$ from
$0.2$ to $1.0$ and hold $\gamma_1 + \gamma_2 = 2$. It shows that the model is
somewhat sensitive to $\gamma_1$ and $\gamma_2$; however, the variation of the
number of topics is mostly less than $10\%$, which is not a major change,
particularly for the average number of topics at levels $2$ and $3$.

In summary, these sensitivity tests indicate that the inferred model is robust
as it tolerates uninformed selections of hyperparameters. The hyperparameters
does have an impact on the learned tree structure, but only in a minor way.  A
specific caution is that one should choose $\gamma_1$ and $\gamma_2$ with more
care than do $\beta$.  Practically, one may compare the perplexities at
different values of $\gamma_1$ and $\gamma_2$, and elect the pair with lower
perplexity.

\subsection{Example RobotStudio Topic Hierarchy}

The result of our approach is a topic hierarchy learned from the combined interaction
and software crash dataset. The tree hierarchy communicates a succinct model of the observed
interactions, where each topic represents a group of commonly co-occurring interactions and
the hierarchy encodes a relationship between general or popular topics and ones that are more
specific and rare.

One may explore the hierarchy either bottom-up or top-down to observe its
structure, or begin with a specific event, such as an exception or stack trace,
and move in both directions with the idea of understanding the context of user
behavior that produces the exception. For instance,
Figure~\ref{fig:user_survey}\subref{subfig-2:robapi} shows a topic hierarchy
learned from the dataset centered on an exception. The hierarchy shows a parent
topic and two of its child topics. Since the messages with dominant
probabilities are about {\em simulation}, we interpret the parent topic to
indicate that developers are starting, stopping, and stepping through a
simulation using RobotStudio.  The two child topics exhibit two
sub-interactions when the user is doing the simulation. The first child,
illustrated immediately below its parent indicates that the user conducts a
conveyor simulation.  The second child indicates that the simulation includes
the user's action that leads the simulated robot {\em moving to} a different
location, which is often accompanied with saving project state, perhaps,
because it is prudent to save the project state before a path change. Thus, we
may conclude that this topic hierarchy suggests that the user starts with a
more generic activity, simulating a robot, and the simulation consists of
multiple sub-interactions.  It also shows that the exception indicated by the
message \code{$\ldots$RobApiException$\ldots$} often occurs with the simulation
for controlling a conveyor.

\subsection{RobotStudio Developer Survey}

In order to assess the interpretability and value of our technique, we
conducted a survey of RobotStudio developers using the model we extracted from
the user interaction dataset of this application. Note that they are the
individuals who develop and maintains RobotStudio, and are not users who use
RobotStudio in production.  One important goal is to help the developers from
using the model built from the data collected from the users.  The survey
consisted of a sample of five random RobotStudio exceptions that we show
to the developers one at a time together with their surrounding context
hierarchy in the survey.

\begin{figure*}[!htbp]
  \centering
    \subfloat[Context hierarchy for FormatException.\label{subfig-1:format}]{%
    \includegraphics[width=0.75\textwidth]{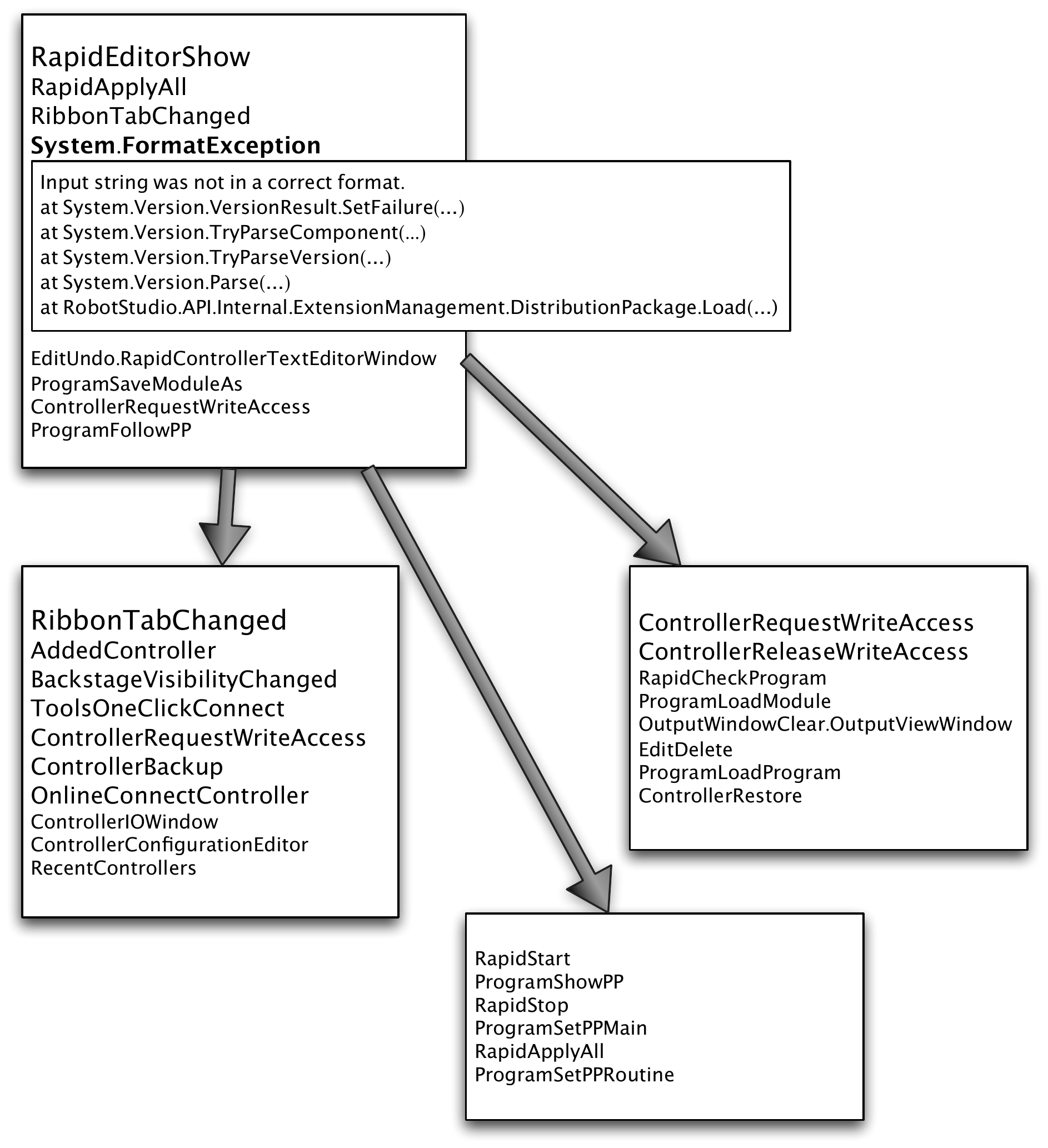}
    }
    %\hfill

    \subfloat[Context hierarchy for RobApiException.\label{subfig-2:robapi}]{%
    \includegraphics[width=0.75\textwidth]{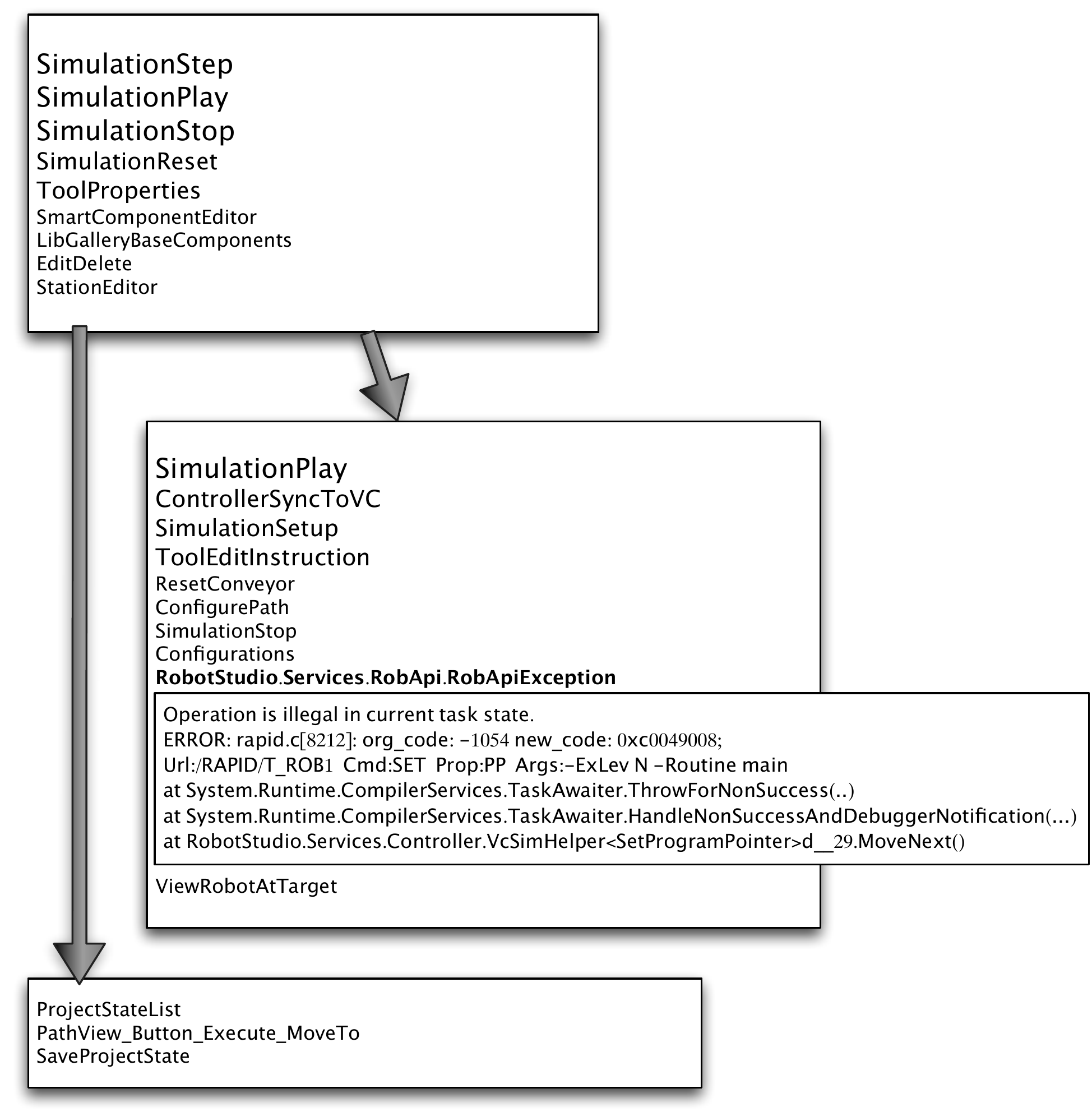}
    }
    \caption{Two of the exception hierarchies presented to RobotStudio developers in survey, where
    font size coarsely approximates the probability of a message in a particular context.
    RobApiException (right) resulted in much higher usefulness ratings by the survey respondents relative
    to all the remaining exception in the survey.}
    \label{fig:user_survey}
\end{figure*}

We sent the composed survey via e-mail to the entire RobotStudio development
team. The team consists of 17 individuals, out of which we received 6
responses. All but one of the respondents had 3 or more years of experience on
the RobotStudio team and all of them had worked as software developers for at
least 3 years. Five out of six respondents were familiar with the RobotStudio
interaction dataset, and had examined it in the past, and all of them believed
that knowing which commands in the interaction log an exception co-occurs with
could be helpful in debugging. Figure~\ref{fig:user_survey} displays two of the
images shown in the survey, which depict an exception and its nearby
surrounding command context hierarchy. Below, we highlight the salient
conclusion from the study, coupled with the evidence to support them, including
any additional relevant explanation extracted from open-ended questions in the
survey.

\vskip 0.3cm

\noindent
\textbf{The model was very useful for understanding and debugging some exceptions, but not
useful for others.} The survey showed a strong variance between the responses
for the usefulness of specific parts of the model and specific exceptions. For instance,
for \code{RobApiException}, listed in Figure~\ref{fig:user_survey}\subref{subfig-2:robapi}, the respondents rated the
{\em usefulness of the usage context in understanding the exception} an average of 7.83
(s = 1.52) on a scale of 1 (least useful) to 10 (most useful). This high rating can be contrasted
to the usefulness rating received by the usage context of the remaining 4 exceptions:
\code{FormatException} - 4.0/10.0 (s = 2.83); \code{ApplicationException} - 3.66/10.0 (s = 3.44);
\code{KeyNotFoundException} - 4.0/10.0 (s = 1.3); \code{GeoException} - 3.83/10.0 (s = 2.92).
Three of the developers already formed the same hypothesis for the fault by examining the model
for \code{RobApiException}, stating the following:\\
{\em [...] VC returns an error saying that we cannot set the program pointer to
main in the current execution state. Perhaps RobotStudio tries to move the
program pointer when it is in running state.}

For the less useful exception models, a number of the RobotStudio developers
suggested a concrete set of improvements that
they believed would raise its level of usefulness, including labeling each of the contexts
and providing additional command characteristics, whenever available,
to make the model clearer. For instance, one participant stated:\\
{\em ``Its like watching the user over the shoulder but too far away. I can see which
tools and windows he or she opens, which commands are issued. But I cannot see any name of an object, no version
number of a controller, no file name, not really anything concrete and specific.
I think that needs to be tied in.''}

Additionally, the survey result that some exceptions are more useful while the
others are not based on the users' ratings may be in part attributed to the
following observation. Some exceptions, e.g., \code{FormatException} and
\code{KeyNotFoundException} may actually not results of program faults
because programmer often use them for input validation\footnote{See the Stack Overflow
discussion ``{\em Is it a good or bad idea throwing Exceptions when validating
data?}'' at
\url{https://stackoverflow.com/questions/1504302/is-it-a-good-or-bad-idea-throwing-exceptions-when-validating-data}
and many other discussions on the subject.}. And yet, when asked about
\code{FormatException}, one developer stated:\\ 
``{\em [...] it tells me that the
user explicitly or implicitly (as far as I remember it is always done
explicitly) was loading a distribution package. The package has it version
number defined as part of the root folder name. The version part of the folder
name could not be parsed to a .NET Version object.}''  \\
In contrast, the
developers view exceptions like \code{RobApiException} and their corresponding
stack traces are more useful because these exceptions are about the movement
and the control of the industrial robot, and perceive them as the results of
actual program faults as discussed above.

%% file: flowchart.tex
\begin{figure}
\centering
\begin{tikzpicture}[scale=1.0,x=0.3in,y=0.3in,z=0.3in]
    \tikzstyle{every node}=[font=\footnotesize]

    \tikzstyle{mstep} = [rectangle, text centered, 
    text width=5em,align=center, rounded corners,
	minimum height=4em,
    draw=black, inner sep=5pt]

    \tikzstyle{connect} = [arrows=->, black]

    \node(s1)[mstep] at (0, 0) {Forming Corpus};
    \node(s2)[mstep] at (4, 0) {Learning Model};
    \node(s3)[mstep] at (8, 0) {Assessing Parameters};
    \node(s4)[mstep] at (12, 0) {Evaluating Model};

    \draw[connect](s1) to (s2);
    \draw[connect](s2) to (s3);
    \draw[connect](s3) to (s4);

\end{tikzpicture}
    \caption{The processing pipeline consists of the following steps, which, although presented linearly, are iterative.
    (a)  {\em Forming Corpus} -- we divide interaction traces into
    sessions, and each session into one or more windows. Scanning the windows, 
    we obtain the vocabulary of
    the corpus. To improve computational efficiency and numerical statbility,
    we remove the words that are overly frequent and those too rare.  
    (b) {\em Learning Model} -- we divide the corpus into the training
    dataset and the testing (held-out) dataset, and start with an initial set
    of parameters to infer a model, and vary these parameters. For each set of
    parameters, we obtain a model.
    (c) {\em Assessing Parameters} -- by computing perplexity on the
    held-out dataset, we determine whether the model converges and
    whether the model is sensitive to parameters, which informs us an
    appropriate set of parameters. Use the parameters, we obtain a final model for
    evaluation. 
    (d) {\em Evaluating Model} -- using a set of randomly selected stack traces
    and their usage contexts, we evaluate the quality of the model by
    analyzing the responses of the developers to the survey.}
\label{fig:pipeline} 
\end{figure}

%% file: sec_threat.tex
\section{Threats to Validity} \label{sec:threat}

This paper presents an exploratory study of using hierarchical topic modeling on large-scale interaction data for 
the purpose of building a hierarchy of usage contexts surrounding stack traces. Such contexts can be useful to
understand or debug software faults that exhibit specific stack traces.  The assumptions embedded in a hierarchical topic model, 
such as, the ``bag of words'' assumption for words in a document, the windowing method, and the modeling approach, are a
source of internal threats to validity of our study. To mitigate this threat, we follow prior established techniques for applying topic models. Also, prior studies have successfully analyzed interaction data using topic models with the ``bag of words'' assumption and a windowing method~\cite{Damevski_Predicting_2017,7515925}.  

In our study, we relied solely on RobotStudio interaction traces to build our model. Therefore, our study’s results may not transfer to other 
interaction traces or platforms. To mitigate this threat we posit that the long timespan and large scale of the Robot Studio interaction
traces, including this development environment's use of extensions that extend its capability, offer a significant amount of
diversity to our technique.

We surveyed RobotStudio developers to evaluate the usefulness of our hierarchy
of contexts. Although the evaluation shows positively that the hierarchy is
helpful to debug software faults, the survey sample size is too small to
provide robust and generalizable conclusions.

The work also suffers from external threats to validity because we surveyed
developers to assess the usefulness of the hierarchy of usage contexts. 
One threat is that the surveyed developers may be prone to offer positive answers as they 
know that we will analyze their responses to the survey, i.e., the observer effect. The other is that our approach may be new to them, and this novelty may influence them to respond positively. To mitigate this threat, we followed standard approaches for creating developer surveys and frequently prompted the survey respondents to specify a rationale for their opinions.

%% file: sec_related.tex
\section{Related Work}\label{sec:related-work}

Although researchers have applied topic models to analyze software engineering
data~\cite{Chen2016, Panichella:2013:EUT:2486788.2486857, 7515925,
Damevski_Predicting_2017}, they have not explored hierarchical topic models, in
particular, Bayesian non-parametric hierarchical topic models that offers
severarl advantages to analyze software engineering data, such as interaction
traces. We focus our related work discussion on the set of prior work that
exists, separately, for both of the data types used in this work, i.e., for
mining and understanding both application crash reports and interaction data.

As interaction data is large-scale, consisting of multiple messages per minute
of user interaction with the application, a common goal is to extract
high-level behaviors from the data that express common behavioral patterns
exhibited by a significant cluster of users. Numerous approaches have been
suggested to extract such behaviors from IDE data, using hidden Markov models,
sequential patterns, Petri nets, and
others~\cite{Damevski:2016:IED:2901739.2901741,
Murphy-Hill:2012:ISD:2393596.2393645, 1316839}, with the purpose of extracting
high-level common behaviors exhibited by developers in the field. Our prior
work explores the use of the Latent Dirichlet Allocation topic modeling
technique, more specifically its temporal variant, for the prediction and
recommendation of IDE commands for a specific
developer~\cite{Damevski_Predicting_2017}.

Mining software crash reports have been a popular area of study in recent
years, with the ubiquity of systems that collect these reports and the
availability of public datasets. Here we highlight only the most relevant
studies, which focus on mining exceptions and stack traces in a corpus of crash
reports.

% debug specific types of bugs, such as, performance bugs etc
Han et al.\@ built wait graphs from stack traces and other messages to diagnose
performance bugs~\cite{Han:2012:PDL:2337223.2337241}. Dang et al.\ clustered
crash reports based on call stack
similarity~\cite{Dang:2012:RMC:2337223.2337364}, while  Wu et al.\ located
bugs by expanding crash stack with functions in static call graphs from crash
reports that contains stack traces~\cite{Wu:2014:CLC:2610384.2610386}.  Davie
et al.\ researched whether a new bug in the same source code as known bug can
be found via bug report similarity measures~\cite{6385108}.

% reduce data & prioritize bug fixes
Crash reports that contains stack traces can be too numerous for engineers to
manage. Dhaliwal et al.\ investigated how to group crash reports based on
bugs~\cite{6080800}.  Kaushik and Tahvildari applied information retrieval
methods or models to detect duplicate bug reports. They compared multiple
information retrieval methods and models including both word-based models and
topic-based models~\cite{6178863}. Williams and Hollingsworth used source code
change history of a software project to drive and help to refine the search for
bugs~\cite{1463230}.

% reuse debugging knowledge
Since bug reports are duplicative and prior knowledge may be used to fix new
bugs, crash reports can help reuse debugging knowledge. Gu et al.\ created a
system to query similar bugs from a bug reports
database~\cite{Gu:2012:RDK:2384616.2384684}.

% difference
Different from prior work, our aim here is to produce a contextual
understanding of stack traces, and their relationship with user interactions.
This is based on a large set of interaction traces with embedded stack traces,
where a stack trace can be considered as a special message in the interaction
traces. While in this paper we always assume a dataset with already combined
interaction and stack traces, they need not be a priori, as long as relatively
reliable timestamps exist in both data sources.  The proposed approach is also
resilient to minor clock synchronization issues that may arise if combining
stack traces and interaction traces that are collected on disparate machines,
since it does not require perfect message ordering.

%% file: sec_conclusion.tex
\section{Conclusions}\label{sec:end}
Large quantities of software interaction traces are gathered from complex
software daily. It is advantageous to leverage such data to improve software
quality by discovering faults, performance bottlenecks, or inefficient user
interface design. We posit that high-level comprehension of these datasets, via
unsupervised approaches to dimension reduction, is useful to improving a myriad
of software engineering activities. In this paper, we aim at modeling a large
set of user interaction data combined with software crash reports. We leverage
a combined dataset collected from ABB RobotStudio a software application with
many thousands of active users.  The described approach is novel in attempting
to model the combination of the two datasets.

As a modeling technique, hierarchical models, such as, the Nested Hierarchical
Dirichlet Process (NHDP) Bayesian non-parametric topic model enable human
interpretation of complex datasets.  The model allows us to extract topics,
i.e., probability distributions of interactions and crashes, from the document
collections and assemble these topics into tree-like structure. The
hierarchical structure of the model allows browsing from a more generic topic
to a more specific topic. The tree also reveals certain structure among users'
interaction with the software. Most importantly, the structure also
demonstrates an understanding how an exception co-occur with other messages,
and thus provide a context on these messages. We surveyed ABB RobotStudio
developers who consistently found parts of the model very useful, although
significant more work is required to understand and predict the parts of the
model that yielded no insight to the developers. The future work also includes
investigating semi-supervised learning models that can leverage developer
feedback in formulating an interpretable and useful model.